\documentclass[reprint, twocolumn,superscriptaddress,
nofootinbib,
amsmath,amssymb, aps, pra,
]{revtex4}

\usepackage{graphicx}
\usepackage{subfigure}
\usepackage{dcolumn}
\usepackage{bm}
\usepackage{color} 
\usepackage{CJK}
\usepackage{dsfont}
\usepackage[extension=xxx]{hyperref}

\newcommand{\beq}{\begin{equation}}
\newcommand{\eeq}{\end{equation}}
\newcommand{\beqa}{\begin{eqnarray}}
\newcommand{\eeqa}{\end{eqnarray}}
\def\ra{\rangle}
\def\beq{\begin{equation}}

\begin{document}
\title{Interferometer with a driven trapped ion}
\author{S. Mart\'\i nez-Garaot}
\affiliation{Department of Physical Chemistry, Universidad del Pa\'{\i}s Vasco - Euskal Herriko Unibertsitatea,
Apdo. 644, Bilbao, Spain}
\author{A. Rodriguez-Prieto}
\affiliation{Departament of Applied Mathematics, Universidad del Pa\'{\i}s Vasco - Euskal Herriko Unibertsitatea, Bilbao, Spain}
\author{J. G. Muga}
\affiliation{Department of Physical Chemistry, Universidad del Pa\'{\i}s Vasco - Euskal Herriko Unibertsitatea,
Apdo. 644, Bilbao, Spain}
\date{\today}

\begin{abstract}
We propose an interferometric  measurement of weak forces using a single ion subjected to 
designed time-dependent spin-dependent forces. Explicit expressions of the relation between the unknown 
force and the final populations are found considering different scenarios, such as the weak force acting on two branches, on one branch, or having errors in the driving force. The flexibility to design the trap trajectories is used to minimize errors due to anharmonicities in the trap. 
The advantages of the approach are the use of geometrical phases, which provides stability, 
the possibility to design faster-than-adiabatic processes with sensitivity control, 
and the independence of the results on the motional states for the small-oscillations regime in which the effective potentials
are purely harmonic.    
  
\end{abstract}

\maketitle

\section{Introduction}
Atom interferometry \cite{Berman1997} works by splitting and recombining the atomic wavefunction, whose interference pattern is sensitive to the 
differential phase accumulated during the separation. It provides impressive sensitivities for inertial sensors and high-precision measurements in gravimeters, gyrometers and velocity sensors. It has been predicted that quantum-enhanced sensors will be available in the market within five years \cite{NSTC2016}. An important open challenge is to miniaturize the interferometers and facilitate their use, for example to measure potential gradients at ultrashort scale and to detect weak forces \cite{Bollinger2017, Bollinger2010,Ivanov2016}, or single-photon scattering events \cite{Roos2013}. Several schemes are currently investigated where wavefunction branches are separated by internal-state dependent potentials, using   
thermal ensembles of cold atoms (rather than condensates to reduce interactions) on chips  \cite{Ammar2015,Dupont-Nivet2016}, or single atoms in optical lattices \cite{Steffen2012}. This ``driven interferometry'' presents similarities with phase gates based on ions in linear traps,  where the trap trajectory is engineered to give a certain chosen phase for each configuration of internal states \cite{Leibfried2003,Palmero2017}.
We explore here this connection in detail, applying inverse engineering techniques used for phase gates \cite{Palmero2017} 
to design an ion interferometer and measure unknown forces. The scheme provides the stability properties of phase gates, namely, the independence of the final phase with respect to motional excitation (temperature), and the geometric character of the phase.  
Moreover, the sensitivity of the interferometer and the process time may be chosen in principle at will,
subjected to technical limitations, to avoid decoherence and visibility 
loss.            
  
Specifically, our setting involves a single ion with two internal states (denoted as ``spin up'', $|\uparrow\rangle$, and ``spin down'', $|\downarrow\ra$) in  harmonic traps.    
The ion state can be written as
\beq
\label{i_s}
\Psi(x,t) = a_{\uparrow} | \uparrow \rangle \psi_{\uparrow}(x,t) + a_{\downarrow}  | \downarrow \rangle
\psi_{\downarrow}(x,t),
\eeq
where $\psi_{\uparrow}(x,t)$ and $\psi_{\downarrow}(x,t)$ are the motional states for the two internal levels, in coordinate representation. 
At time zero $\psi_{\uparrow}(x,0)=\psi_{\downarrow}(x,0)$. The states are driven by 
spin-dependent forces so that the modulus of 
 $|\langle \psi_{\uparrow}(t_f)|\psi_{\downarrow}(t_f)\rangle|$ is one (or nearly one due to errors or unknown forces),
at a final time $t_f$.   
The measurement of the phase is done through measurements of populations. 
A common method is to apply a $\pi/2$ pulse \cite{Leibfried2003} 
\beqa
\label{pipulse}
| \uparrow \rangle & \rightarrow & \frac{1}{\sqrt{2}} (| \uparrow \rangle - | \downarrow \rangle), 
\nonumber \\
| \downarrow \rangle & \rightarrow & \frac{1}{\sqrt{2}} (| \uparrow \rangle + | \downarrow \rangle),
\eeqa
where the electromagnetic field phase has been fixed to $\pi/2$ \cite{Berman1997}.
Substituting (\ref{pipulse}) in Eq. (\ref{i_s}) at time $t_f$, the  state becomes
\beqa
\Psi(x,t_f) &=& \frac{1}{\sqrt{2}} \left [ a_{\uparrow}\psi_{\uparrow}(x,t_f) +  a_{\downarrow}
 \psi_{\downarrow}(x,t_f) \right ] | \uparrow \rangle 
\nonumber\\
&-& \frac{1}{\sqrt{2}} \left [ a_{\uparrow}\psi_{\uparrow}(x,t_f) - 
a_{\downarrow}\psi_{\downarrow}(x,t_f) \right ] | \downarrow \rangle,
\eeqa
and the population of each spin configuration will be
\beqa
\label{populations}
&&\hspace{-.5cm}P_{\uparrow}(t_f)=\frac{1}{2} (|a_\uparrow|^2\!+\!|a_\downarrow|^2)\!+\!\Re{\rm e}\! \left [ a_\downarrow^*a_\uparrow \langle \psi_{\downarrow}(t_f)|\psi_{\uparrow}(t_f) \rangle \right ]\!,
\nonumber \\
&&\hspace{-.5cm}P_{\downarrow}(t_f)=\!\frac{1}{2} (|a_\uparrow|^2\!+\!|a_\downarrow|^2)\!-\!\Re{\rm e}\! 
\left [a_\downarrow^*a_\uparrow \langle \psi_{\downarrow}(t_f)|\psi_{\uparrow}(t_f) \rangle \right ]\!.
\eeqa
We consider the simplest case $a_\uparrow=a_\downarrow=\frac{1}{\sqrt{2}}$, so 
%
\beqa
\label{populations_2}
P_{\uparrow}(t_f)&=&  \frac{1}{2}+ \frac{1}{2}\Re{\rm e} \left [ \langle \psi_{\downarrow}(t_f)|\psi_{\uparrow}(t_f) \rangle \right ], 
\nonumber \\
P_{\downarrow}(t_f)&=&\frac{1}{2}-\frac{1}{2}\Re{\rm e} \left [\langle \psi_{\downarrow}(t_f)|\psi_{\uparrow}(t_f) \rangle \right ],
\eeqa
and the overlap can be written as 
\beq
\label{o}
\langle \psi_{\downarrow}(t_f)|\psi_{\uparrow}(t_f) \rangle=e^{i\Delta\phi(t_f)}|\langle \psi_{\downarrow}(t_f)
|\psi_{\uparrow}(t_f) \rangle|. 
\eeq
If the modulus in Eq. (\ref{o}) is indeed $1$, the interference pattern of the populations oscillates with $\Delta\phi(t)$ with maximal visibility, otherwise the visibility is reduced. 
An additional, unknown, small  force will affect the phase difference as well as the modulus. The effect on the phase difference will produce a shift in the oscillation of the interference pattern while the effect on the modulus will decrease the visibility.  

In Section \ref{br} we will analyze the phases accumulated along the branches. 
Section \ref{model} describes 
how 
invariant based engineering allows us to control the process-time without residual excitations. 
It also provides a frame to calculate corrections. 
In Section \ref{effect_c} we study the effect of a homogeneous, small, constant and unknown offset force $c$, by means of the Lewis-Riesenfeld invariants of motion. First we consider   that both branches of the interferometer are subjected to such a force $c$, analyzing also the effect of a constant error $\epsilon$ in the driving force. Finally, we examine the case in which only one of the branches is perturbed by $c$. 
In all scenarios studied explicit expressions are found for the final phase and visibility.   
\section{Phases and forces\label{br}}
Consider a single positive ion of charge $e$ and mass $m$, trapped in a radially-tight, effectively one-dimensional (1D) trap along $x$ with angular frequency $\omega$.
A spin-independent homogeneous, constant force $c$ that we want to measure 
is applied to the trapped ion in the longitudinal direction.
We assume that the ion may be treated as a two-level system  
affected by additional ``spin-dependent'' forces, 
opposite for the two internal levels,  $f(t;\sigma^z)=\sigma^zf(t)$. 
Other cases may as well be considered as in \cite{Palmero2017}.  
Here $\sigma^{z}=\pm 1$ are eigenvalues of the Pauli matrix $\sigma^{z}$ 
for spin up ($\sigma^{z}=+ 1$) and spin down ($\sigma^{z}=-1$) internal states, respectively. 
Off-resonant lasers induce the spin-dependent forces that are assumed to be homogeneous over the extent of 
the motional state (Lamb-Dicke regime).

\begin{figure}[h!]
{\includegraphics[width=85mm]{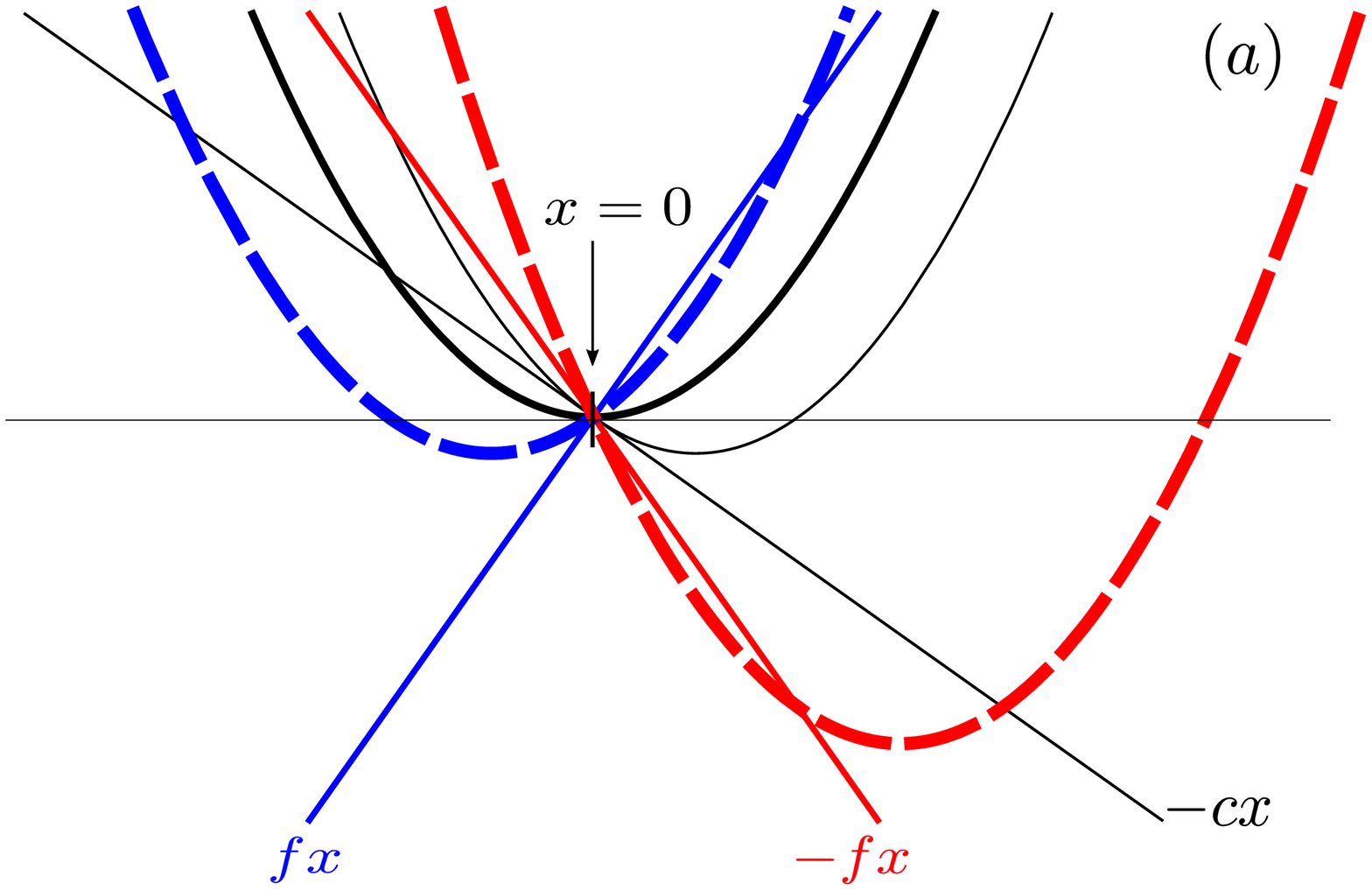}}
{\includegraphics[width=85mm]{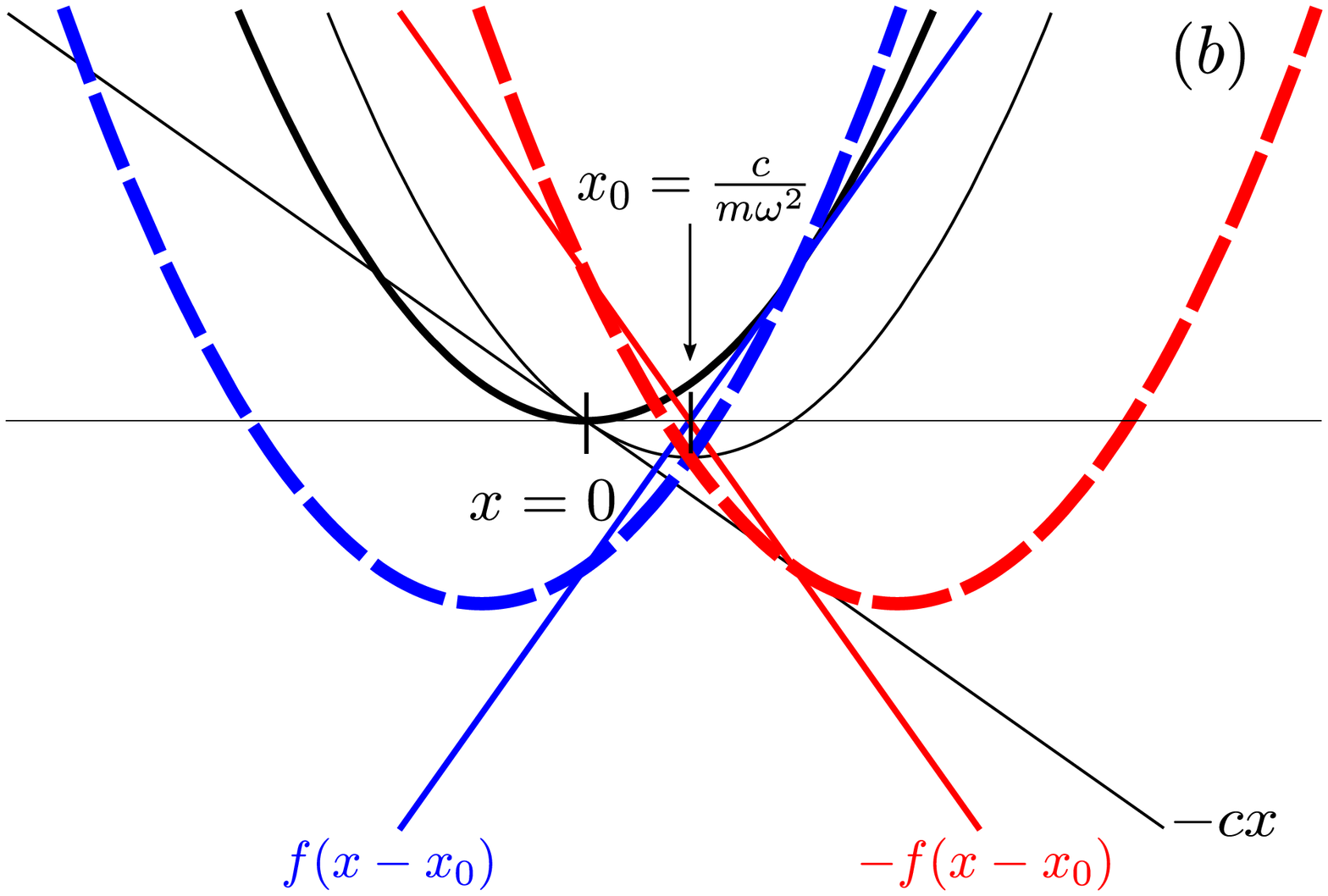}}
\caption{(Color online) Schematic configurations of potentials for spin-down (left, dashed blue line) 
and spin-up (right, dashed red line) components at some intermediate time $t$.
The potentials are formed by summing a harmonic 
term centered at $x=0$ (thick black solid line) with angular frequency $\omega$,
a potential $-cx$ (thin black solid straight line) and spin-dependent 
linear potentials for spin-down (solid blue line) and spin-up (solid red line) components. The crossing-point
of these spin-dependent 
linear potentials is at $x=0$ in (a) and in $c/(m\omega^2)$ in (b). $c/(m\omega^2)$
is the location of the potential minimum of the harmonic potential (thin black solid line) displaced by the force $c$.  
The spin-dependent harmonic potentials have different heights  in (a). The difference, integrated over the process time,  generates a non-zero interferometric phase. On the contrary, when the spin-dependent 
linear potentials cross at $c/(m\omega^2)$,  the potential heights are equal throughout the process 
so the phase is zero (b).    
}
\label{fig:scheme_potentials}
\end{figure}

The Hamiltonian of such a system can be written as 
\begin{eqnarray}
\label{DidiH_A}
H&=&\frac{p^{2}}{2m}+\frac{1}{2}m\omega^{2}x^{2}-cx-f(t,\sigma^{z})\left[x-x_{0}(t)\right]
\nonumber\\
&=&\frac{p^{2}}{2m}+\frac{1}{2}m\omega^{2}\left(x-\frac{c}{m\omega^{2}}\right)^{2}-\frac{c^{2}}{2m\omega^{2}}
\nonumber\\
&-&f(t,\sigma^{z})\left[x-x_{0}(t)\right],
\end{eqnarray}
where $x_0(t)$ may depend on time. Note the role of $x_0(t)$ as ``crossing point'' of the potential energies for the spin-dependent forces (see Fig. \ref{fig:scheme_potentials}).  
Introducing the new variable $\tilde{x}=x-\frac{c}{m\omega^{2}}$ we may rewrite the Hamiltonian as 
\begin{eqnarray}
\label{DidiH_A_new}
H&=&\frac{p^{2}}{2m}+\frac{1}{2}m\omega^{2}\tilde{x}^{2}
\nonumber\\
&-&f(t,\sigma^{z})\left[ \tilde{x}-x_{0}(t)+ \frac{c}{m\omega^{2}}\right]-\frac{c^{2}}{2m\omega^{2}}
\nonumber\\
&=&\frac{p^{2}}{2m}+\frac{1}{2}m\omega^{2}\tilde{x}^{2}-f(t,\sigma^{z})\tilde{x}
\nonumber\\
&-&\frac{c}{2m\omega^{2}}\left[c+2f(t,\sigma^{z})\right]+f(t,\sigma^{z})x_{0}(t).
\end{eqnarray}
Let us now consider separately the spin-up and spin-down branches.

For the spin-up branch the spin-dependent force is $f(t,\sigma^{z})=f(t)$, and the Hamiltonian reads 
\begin{equation}
H^{\uparrow}=\frac{p^{2}}{2m}+\frac{1}{2}m\omega^{2}\tilde{x}^{2}-f(t)\tilde{x}
-\frac{c}{2m\omega^{2}}\!\left[c+2f(t)\right]+f(t)x_{0}(t).
\end{equation}
Separating the effect of purely time-dependent terms with phase factors,
the solution of the Schr\"{o}dinger equation for this Hamiltonian is
\begin{eqnarray}
\label{Didi_Phi_A_up}
\psi_{\uparrow}^{c\neq 0}(\tilde{x},t)  &=& e^{\frac{i}{\hbar}\frac{c}{2m\omega^{2}}
\int_{0}^{t} \left[c+2f(t')\right]dt'}
\nonumber\\ 
&\times& e^{-\frac{i}{\hbar}\int_{0}^{t} x_{0}(t')f(t')dt'}  \psi_{\uparrow}^{c=x_0=0}(\tilde{x},t),
\end{eqnarray}
where $\psi_{\uparrow}^{c=x_0=0}(\tilde{x},t)$ is the solution
of the Schr\"{o}dinger equation for the system whose Hamiltonian is $H^{\uparrow}$ with $c=0$ and $x_0=0$.   

The phase accumulated when traveling through the spin-up branch with respect to 
$\psi_{\uparrow}^{c=x_0=0}(\tilde{x},t)$ is  therefore
\beqa
\label{Didi_Phase_A_up}
\phi_{\uparrow}(t)=\frac{c}{2\hbar m \omega^{2}}\! \int_{0}^{t}
\!\left[c+2f(t')\right]dt'-\frac{1}{\hbar}\int_{0}^{t}x_{0}(t')f(t')dt'.
\nonumber \\
\eeqa
The driving force $f(t)$ is designed inversely from the Newton equation
\begin{equation}
\label{DidiNewton}
\ddot  y(t)+\omega^{2}y(t)=\frac{f(t)}{m}, 
\end{equation}
for particular solutions $y(t)=\alpha(t)$ that satisfy the boundary conditions 
\beq
\alpha(t_{b})=\dot \alpha(t_{b})=\ddot \alpha(t_b)=0
\label{bc_y}
\eeq
at the boundary times $t_{b}=0, t_{f}$. Here and throughout the paper dots denote time derivatives. We shall distinguish forces designed this way with the subscript $\alpha$, 
i.e., 
\begin{equation}
\ddot  \alpha(t)+\omega^{2}\alpha(t)=\frac{f_\alpha(t)}{m}.
\label{fal}
\end{equation}
The $f_\alpha(t)$ vanish at the boundary times. Moreover,  
\begin{eqnarray}
\label{Integral}
\int_{0}^{t_{f}}f_\alpha(t)dt&=&m\int_{0}^{t_{f}}\left[\ddot  \alpha (t)+\omega^{2}\alpha(t) \right]dt=
m\left[\dot\alpha(t)\right]_{0}^{t_{f}}
\nonumber \\
&+&m\omega^{2}\!\!\int_{0}^{t_{f}}\!\!\alpha(t)dt=m\omega^{2}\!\!\int_{0}^{t_{f}}\!\!\alpha(t)dt.
\label{integr}
\end{eqnarray}
Then we may write the phase accumulated at final time as  
\begin{equation}
\label{Didi_Phase_A_up_new}
\phi_{\uparrow}(t_{f})=\frac{c^{2}t_{f}}{2\hbar m \omega^{2}}
+\frac{c}{\hbar} \int_{0}^{t_{f}}\!\alpha(t)dt-\frac{1}{\hbar}\int_{0}^{t_{f}}\!x_{0}(t)f_\alpha(t)dt.
\end{equation}

Similarly, for the spin-down branch $f(t,\sigma^{z})=-f(t)$, and 
\begin{eqnarray}
\label{Didi_Phi_A_down}
\psi_{\downarrow}^{c\neq 0}(\tilde{x},t) &=& e^{\frac{i}{\hbar}\frac{c}{2m\omega^{2}}  \int_{0}^{t} \left[c-2f(t')\right]dt'}
\nonumber\\ 
&\times& e^{\frac{i}{\hbar}\int_{0}^{t} x_{0}(t')f(t')dt'}  \psi_{\downarrow}^{c=x_0=0}(\tilde{x},t),
\end{eqnarray}
so the phase accumulated with respect to $\psi_{\downarrow}^{c=x_0=0}(\tilde{x},t_f)$
for the special forces $-f_\alpha$ is
\begin{equation}
\label{Didi_Phase_A_down_new}
\phi_{\downarrow}(t_{f})=\frac{c^{2}t_{f}}{2\hbar m \omega^{2}}-\frac{c}{\hbar}
\int_{0}^{t_{f}}\!\alpha(t)dt+\frac{1}{\hbar}\!\int_{0}^{t_{f}}\!x_{0}(t)f_\alpha(t)dt.
\end{equation}
We assume that $\psi_{\downarrow}^{c=x_0=0}(\tilde{x},0) =
\psi_{\uparrow}^{c=0,x_0=0}(\tilde{x},0)$, and that this common initial state can be arbitrary. 
In the next section we shall see that for the forces $\pm f_\alpha$, 
then $\psi_{\downarrow}^{c=x_0=0}(\tilde{x},t_f)=\psi_{\uparrow}^{c=x_0=0}(\tilde{x},t_f)$. 
Thus, the phase difference at final time is  
\begin{equation}
\label{Didi_PhaseDifference}
\Delta \phi(t_{f})=\phi_{\uparrow}-\phi_{\downarrow}=
\frac{2c}{\hbar}\int_{0}^{t_{f}}\! \alpha(t)dt-\frac{2}{\hbar} \int_{0}^{t_{f}}\! x_{0}(t) f_\alpha(t)dt.
\end{equation}
This differential phase has two alternative simple readings which we find useful for an intuitive grasp of the 
interference effect. The first one uses the different energy paths of the two traps:   
 $\Delta \phi(t_f)=-\int_0^{t_f} \Delta E(t)/\hbar$, where $\Delta E(t)=E_{min}^\uparrow
-E_{min}^\downarrow$ is the energy shift between the minima of the spin-dependent harmonic potentials (see Fig. \ref{fig:scheme_potentials}); the second one reads $\Delta \phi(t_f)$ geometrically as the difference of phase-space areas covered by the trajectories perturbed by $c$ in a rotating frame (see Fig. \ref{fig:scheme_areas} and 
more details in Appendices A and B). These may be classical trajectories or equivalently trajectories of a wave packet center. 

For $x_0$ constant in time, applying Eq. (\ref{integr}) we have
\begin{equation}
\label{Didi_PhaseDifferenceB_new}
\Delta \phi(t_{f})=\phi_{\uparrow}-\phi_{\downarrow}
=\frac{2}{\hbar}\left[c-m\omega^{2}x_{0}\right] \int_{0}^{t_{f}} \alpha(t)dt.
\end{equation}
In particular, if  $x_{0}=0$ the phase difference is
\begin{equation}
\label{pd0}
\Delta \phi(t_{f})|_{_{x_0=0}}=\phi_{\uparrow}-\phi_{\downarrow}=\frac{2c}{\hbar}\int_{0}^{t_{f}} \alpha(t)dt,
\end{equation}
see Fig. \ref{fig:scheme_potentials} (a), but no phase difference arises if $x_{0}=\frac{c}{m\omega^{2}}$, 
\begin{equation}
\label{Didi_PhaseDifferenceB2}
\Delta \phi(t_{f})|_{_{x_0=\frac{c}{m\omega^{2}}}}=\phi_{\uparrow}-\phi_{\downarrow}=0, 
\end{equation}
since then the energies of the two traps are equal at all times, see Fig. \ref{fig:scheme_potentials} (b). Alternatively, 
Figure \ref{fig:scheme_areas} demonstrates the geometric perspective.  
In the first case, $x_0=0$, the phase-space areas enclosed by the two spin states are not the same so a differential phase shift (proporcional to the difference in areas) arises, see Fig. \ref{fig:scheme_areas} (a). 
Instead, in the second scenario, $x_{0}=c/m\omega^{2}$, the phase-space area is the same for both spin sates so no differential phase shift comes about, see Fig. \ref{fig:scheme_areas} (b). 
For a given choice of $x_0$, the right or left branch areas are equal for all trajectories, irrespective of their initial location, compare solid and dashed line trajectories in Fig. \ref{fig:scheme_areas} (and see Appendices A and B for further details). 
%
%
%
%
%
%
\begin{figure}[h!]
{\includegraphics[width=85mm]{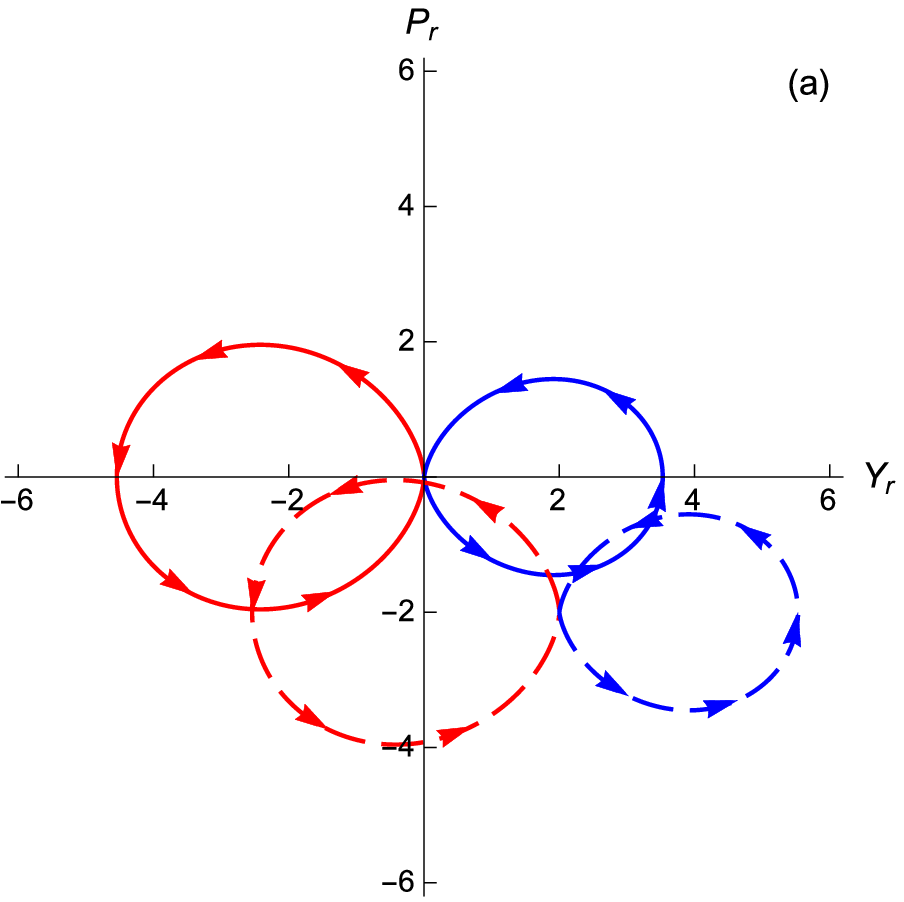}}
{\includegraphics[width=85mm]{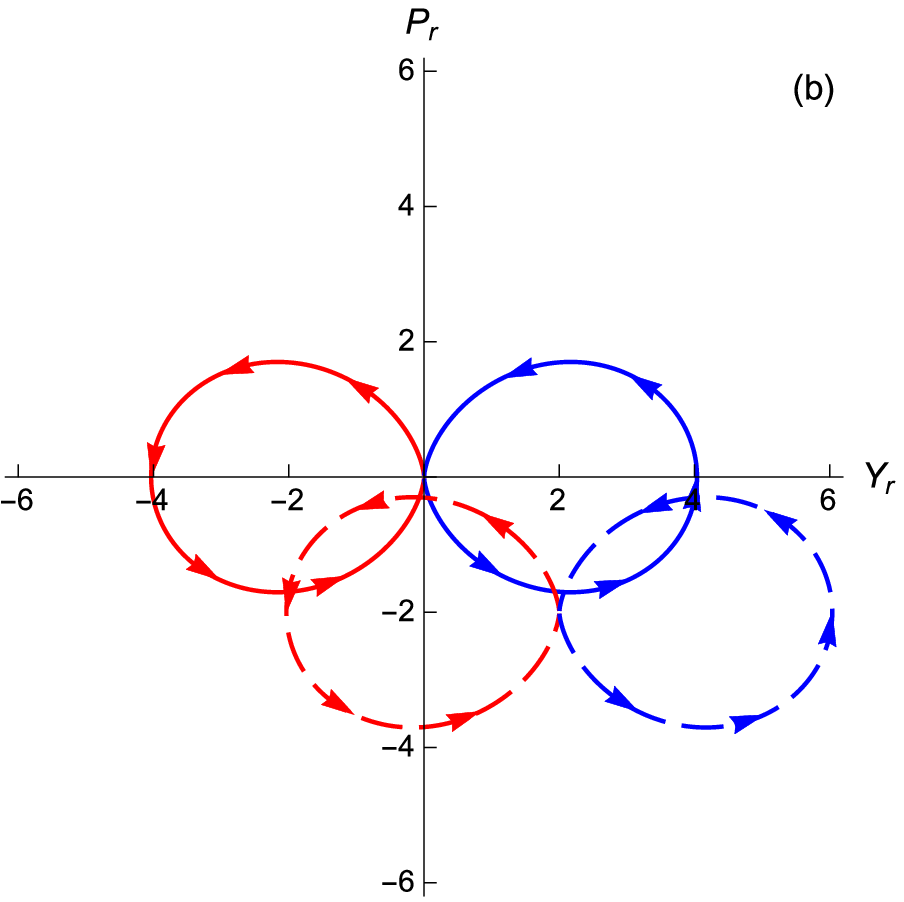}}
\caption{(Color online) Phase space trajectories in spin-dependent potentials [spin-down (right, blue) and spin-up (left, red)] in the rotating frame defined as $Y_r=\Re{\rm e}(e^{i\omega t}Z^{\uparrow,\downarrow})$ and 
$P_r=\Im{\rm m}(e^{i\omega t}Z^{\uparrow,\downarrow})$, where $Z^{\uparrow,\downarrow}=Y+iP$, $Y=\sqrt{\frac{m \omega}{2 \hbar}} y^{\uparrow,\downarrow}$ and 
$P=\sqrt{\frac{m}{2 \hbar \omega}}\dot y^{\uparrow,\downarrow}$.  $y^{\uparrow},y^{\downarrow}$ are solutions of the forced harmonic 
oscillators with forces $f_\alpha+\tilde{c}$, and $-f_\alpha+\tilde{c}$, respectively.  $\tilde{c}=c-m\omega^2 x_0$. 
(a) When $x_0=0$ the areas enclosed by left and right branches are different. (b) 
If $x_0=\frac{c}{m \omega^2}$ the area is the same for both branches. 
The area does not depend on the initial condition $Z(0)$, compare the trajectories for $Z(0)=0$ (solid lines) or  $Z(0)\neq0$ (dashed lines) (see Appendices A and B for more details).    
}
\label{fig:scheme_areas}
\end{figure}
%
%

The physical interpretation of these results is key to measure an 
unknown $c$ with the interferometer. Suppose that $c$ is not only unknown but also acts permanently, 
during the experiment, even before the state is initialized at time $t=0$.      
For $t\le 0$ the minimum of the trap is shifted by the force $c$  from $x=0$ to the point $\frac{c}{m\omega^{2}}$,  
which becomes the center of a stationary wavefunction prepared by cooling the ion. 
The spin-dependent potential terms will be naturally centered so that they cross at that point, i.e., $x_0=\frac{c}{m\omega^{2}}$, and no differential phase may arise. 
The setting is different if, instead, the force $c$ does not act before $t=0$, so that $x_0=0$.
Even if $c$ is acting along $x$, its effect may be turned on by rotating the trap to $x$ from a perpendicular direction. 
In the following we shall assume $x_0=0$ for simplicity unless stated otherwise.
\section{Invariant based engineering}
\label{model}
The configuration with  $c=0, x_0=0$ played a central role in the previous section. 
In this section we shall learn more about this special configuration making use of 
Lewis-Riesenfeld invariants of motion \cite{Lewis1969}, see \cite{Torrontegui2011,Torrontegui2013}. 
The invariants help us to inverse engineer the trap trajectories so that the final states 
meet again at a chosen final time without residual excitation at their original positions.  
This provides not only process-time control but also maximum visibility. 
The invariant formalism  also produces analytical wavefunctions and exact expressions for second order corrections 
(and beyond if desired) of the phase difference and modulus of the overlap due to different errors. 

A dynamical invariant for a Hamiltonian $H$ satisfies 
\beq
\label{i_condition}
\frac{dI}{dt}\equiv \frac{\partial I}{\partial t}+\frac{1}{i \hbar} [I,H]=0.
\eeq
In particular, let $H$ correspond to a forced harmonic oscillator,
\beq
\label{Hamiltonian}
H = \frac{p^2}{2m} + \frac{1}{2} m\omega^2x^2 - F(t)x.
\eeq
We shall later consider different forces $F(t)$. 
  
For such a Hamiltonian, there are quadratic invariants of the form 
\beq
I(t)=\frac{1}{2m} (p-m\dot y)^2+ \frac{1}{2}m \omega ^2 (x-y)^2,
\eeq
where $y(t)$ can be interpreted as a ``reference trajectory'' that satisfies the differential (Newton) equation
\beq
\label{NE}
\ddot y + \omega^2 y =\frac{F(t)}{m},
\eeq
in the forced harmonic potential \cite{Torrontegui2011}.
In the following we shall systematically use reference trajectories $y_0(0)$ 
characterized by the boundary conditions $y_0(0)=0, \dot{y}_0(0)=0$.  
Note that  $\alpha(t)$ in the previous section is a particular case for $F=f_\alpha$.    
The invariant is Hermitian, and has a complete set of eigenstates. Solving
\beq
I(t)\psi_n (x,t)=\lambda_n \psi_n (x,t)
\eeq
we get the time-independent eigenvalues
\beq
\lambda_n=\hbar \omega \left(n+\frac{1}{2}\right),
\eeq
and the time-dependent eigenvectors
\beq
\label{e-v}
\psi_n(x,t)=e^{\frac{i m}{\hbar} \dot y x} \phi_n(x-y), 
\eeq
where $\phi_n(x)$ is the $n$th eigenvector of the stationary oscillator, 
\beq
\label{e-v_o}
\phi_n(x)=\frac{1}{\sqrt{2^n n!}} \left ( \frac{m\omega}{\pi \hbar} \right )^{1/4}
e^{\frac{-m\omega x^2}{2 \hbar}} H_n \left ( \sqrt{\frac{m\omega}{\hbar}} x \right ),
\eeq
and the $H_n$ are Hermite polynomials. 
  
The Lewis-Riesenfeld phases $\theta_n(t)$ must satisfy
\beq
\hbar \frac{d \theta_n}{dt}= \langle \psi_n | i\hbar \frac{\partial}{\partial t}-H | \psi_n \rangle,
\eeq
so that the functions $e^{i\theta_n(t)}\psi_n(x,t)$ are solutions of the time-dependent Schr\"odinger equation.
They form a complete orthogonal basis of ``dynamical modes''. Each of them is centered at the trajectory $y(t)$
[or $\alpha(t)$ for a force $F=f_\alpha$, see Eqs. (\ref{bc_y},\ref{fal})].  

Using Eq. (\ref{e-v}), the Lewis-Riesenfeld phases are given by
\beqa
\label{phases}
\theta_n (t)&=& -\frac{1}{\hbar} \int_0^t dt' \left [\lambda_n + \frac{m}{2}(\dot y^2-\omega^2 y^2) \right ]
\nonumber \\
&=&-\left (n+\frac{1}{2} \right)\omega t -G(t),
\eeqa
where 
\beq
\label{IPphases}
G(t)=\frac{m}{2 \hbar} \int_0^t dt' (\dot y^2-\omega^2 y^2).
\eeq
The solution of the Schr\"odinger equation for the Hamiltonian (\ref{Hamiltonian}) with force $F(t)$, 
can generally be  written as 
\beq
\label{SE_solution}
\psi(x,t)=\sum_n c_n e^{i \theta_n(t)}\psi_n(x,t).
\eeq
It is often  useful to consider a rotating frame
defined as $\psi_I(t)=e^{iH_{0}t/\hbar}\psi(t)$, 
where $H_{0}\equiv p^2/(2m)+m\omega^2 x^2/2$,
so that
\beq
\psi_I(x,t)=e^{-iG(t)}\sum_n c_n \psi_n(x,t)=\sum_n c_n \Psi_n(x,t),
\eeq
where
\beq
\Psi_n(x,t)\equiv e^{-iG(t)} \psi_n(x,t).
\eeq
Note that the phase $\phi(t)=-G(t)$ is the same for any state. 
Its relation to dynamical, geometric phases and phase space-areas is discussed in Appendices A and B.

Let us now assume the spin-up force $F(t)=f_\alpha(t)$ with $c=x_0=0$.   
This force guarantees that all dynamical modes $e^{i \theta_n(t)}\psi_n(x,t)$
end up at the original positions and at rest, 
\beq
e^{i \theta_n(t_f)}\phi_n(x).
\eeq
Integrating $G(t)$ in (\ref{IPphases}) by parts, and using Eqs. (\ref{bc_y}) and (\ref{fal}), 
the phase factor common to all $n$ at final time takes the form
\beq
G_\alpha(t_f)=-\frac{1}{2 \hbar} \int_0^{t_f} dt f_\alpha \alpha,
\label{Ga}
\eeq
where the subscript emphasizes that the special force $f_\alpha$ is 
assumed.  

For the corresponding spin-down branch both the force and the special trajectory change sign, $F=-f_\alpha$,  $\alpha\to -\alpha$,
so the same common phase factor 
$G_\alpha(t_f)$ results at final time. 
Similarly, the $n$-dependent term of $\theta_n(t_f)$ does not change by the force
reflection $f_\alpha\to -f_\alpha$,
so that, for any (common) initial state 
$\psi^{c=x_0=0}_\downarrow(x,0)=\psi^{c=x_0=0}_\uparrow(x,0)$, we have as well $\psi^{c=x_0=0}_\downarrow(x,t_f)=\psi^{c=x_0=0}_\uparrow(x,t_f)$. 
This is a result announced in the previous section.  
\section{Effect of a homogeneous force $c$}
\label{effect_c}
If a homogeneous force $c$ acts, and the spin-dependent forces $\pm f_\alpha$ are applied with no errors to both branches, 
the visibility is one, and the differential phase is found easily as in Section II, see Eq. (\ref{pd0}), for $x_0=0$.
An alternative route leading to the same result makes use of the phases found in Sec. \ref{model} for the 
perturbed harmonic oscillators and trajectories. This methodology is made explicit in the following subsections.   
\subsection{Constant error in the driving force $f$\label{4A}}
To test the stability of the method we consider now an error $\epsilon$ in the spin-dependent driving force besides the effect of the homogeneous one $c$. Therefore, the effective forces acting on both branches are  $f_\alpha+\epsilon+c$ for $\sigma^z=1$ and $-f_\alpha-\epsilon+c$ for $\sigma^z=-1$. 
The Newton equation (\ref{NE}) becomes for the two cases 
\beqa
\label{NE_lr_2}
\ddot y ^\uparrow + \omega^2 y^\uparrow &=&\frac{1}{m}(f_\alpha+\epsilon+c),
\nonumber \\
\ddot y ^\downarrow + \omega^2 y^\downarrow &=&\frac{1}{m}(-f_\alpha-\epsilon+c),
\eeqa
where we consider solutions of the form 
\beqa
\label{sol_per_2}
y_0^\uparrow&=&\alpha + {\delta \alpha}_\epsilon +{\delta \alpha}_c,
\nonumber \\
y_0^\downarrow&=&-\alpha- {\delta \alpha}_\epsilon + {\delta \alpha}_c.
\eeqa
The subscripts $\epsilon$ and $c$ distinguish between the deviations due to the homogeneous force and the ones due to the error. 
The subscript $0$ is a reminder of the initial boundary condition chosen, $y_0(0)=\dot{y}_0(0)=0$.   
It is convenient to use dimensionless versions of these positions and the corresponding momenta, see Eq. (\ref{cuadrature}), 
and take them as real and imaginary parts of dimensionless complex variables $Z_0^\uparrow$ and  $Z_0^\downarrow$. 
We may decompose the two motions as   
\beqa
Z_0^\uparrow(t)&=&Z_{\alpha}(t)+\delta Z_{\epsilon}(t)+\delta Z_{c}(t), 
\nonumber \\
Z_0^\downarrow(t)&=&-Z_{\alpha}(t)-\delta Z_{\epsilon}(t)+\delta Z_{c}(t),
\eeqa
where
\beqa
\label{deltaz_ce}
\delta Z_{\gamma}(t)&=&=e^{-i \omega t} \frac{i\gamma}{\sqrt{2 \hbar m \omega}} \int_0^t d\tau e^{i \omega \tau}
\nonumber\\
&=&\frac{\gamma}{\sqrt{2 \hbar m \omega^3}}(1-e^{-i\omega t}),
\eeqa
for $\gamma=c,\epsilon$, 
and $Z_{\alpha}$ is the dimensionless complex trajectory for $\alpha$.   
Using the expressions for the deviations, 
\beqa
\label{de_dc}
{\delta \alpha_\gamma}&=&\frac{\gamma}{m \omega^2}[1-\cos(\omega t)],
\eeqa
and Eqs. (\ref{zero}) and (\ref{integr}) to do the integrals,  
the phases of the branches (\ref{IPphases}) at final time are
\beqa
&&\hspace{-.5cm}G^\uparrow(t_f)\!=\! \frac{1}{2\hbar}\!\left[m(\dot \delta \alpha_{ \epsilon} \delta \alpha_{\epsilon}\! +\! \dot \delta \alpha_\epsilon \delta \alpha_c\!+\! \dot \delta \alpha_c \delta \alpha_\epsilon\! +\! \dot \delta \alpha_c \delta \alpha_c)_{t_f} \right. 
\nonumber \\
&&\hspace{-.5cm}\left.-\int_0^{t_f}\!\! \alpha f_\alpha dt - 2(c+\epsilon)\! \int_0^{t_f}\!\! \alpha dt - (c+\epsilon)\! \int_0^{t_f}\!\! (\delta \alpha_\epsilon+\delta \alpha_c) dt  \right],
\nonumber \\
&&\hspace{-.5cm}G^\downarrow(t_f)\!=\! \frac{1}{2\hbar}\!\left[ m( \dot \delta \alpha_\epsilon \delta \alpha_\epsilon\! -\! \dot \delta \alpha_\epsilon \delta  \alpha_c\!-\!m \dot \delta \alpha_c \delta \alpha_\epsilon\! +\! \dot \delta \alpha_c \delta \alpha_c)_{t_f} \right.
\nonumber \\
&&\hspace{-.5cm}\left.-\int_0^{t_f}\!\! \alpha f_\alpha dt + 2(c-\epsilon)\! \int_0^{t_f}\! \!\alpha dt + (c-\epsilon)\! \int_0^{t_f}\! \!(\delta \alpha_\epsilon-\delta \alpha_c) dt  \right]\!,
\nonumber\\
\eeqa  
where $(...)_{t_f}$ is a shorthand for calculating $(...)$ at $t_f$.  

The overlap for the mode components is finally
\beqa
&&\hspace*{-.5cm}\langle \Psi_{n}^{\downarrow}(t_f)|\Psi_{n'}^{\uparrow}(t_f)\rangle
= e^{i[G^\downarrow(t_f)-G^\uparrow(t_f)]} \int\! dx e^{\frac{2im}{\hbar} (\dot \delta \alpha_\epsilon)_{t_f} x}\nonumber\\
&&\hspace{-.5cm}\times \phi_{n} [x\!-\!(\delta \alpha_c\!-\!\delta \alpha_\epsilon)_{t_f}] \phi_{n'} [x\!-\!(\delta \alpha_c\!+\!\delta \alpha_\epsilon)_{t_f}].
\eeqa
%
For $n'\geqslant n$, 
\beqa
\label{overlap_epsilon}
&&\hspace{-.7cm}\langle \Psi_{n}^{\downarrow}(t_f)|\Psi_{n'}^{\uparrow}(t_f)\rangle
\!=\! \langle \Psi^{\downarrow}_0(t_f)|\Psi^{\uparrow}_0(t_f)\! \rangle ({\delta Z_\epsilon}(t_f)^*)^{(\!n'\!-n)} 2^{(\!n'\!-n)}
\nonumber \\
&&\hspace*{-.7cm}\times\sum_{k=0}^\infty (-1)^{(k+n'-n)} \frac{n!}{(n-k)!} \sqrt{\frac{n'!}{n!}} \frac{\left [4|{\delta Z_\epsilon}(t_f)|^2 \right]^k }{k! (k+n'-n)!}
\nonumber\\
&&\hspace{-.7cm}= \langle \Psi^{\downarrow}_0(t_f)|\Psi^{\uparrow}_0(t_f) \rangle ({\delta Z_\epsilon(t_f)}^*)^{(n'-n)} \frac{(-2)^{(n'-n)}}{(n'-n)!} \sqrt{\frac{n'!}{n!}} \nonumber \\
&&\hspace{-.7cm}\times_1F_1 \left\{-n; n'-n+1; 4|{\delta Z_\epsilon}(t_f)|^2 \right \},
\eeqa
where $_1F_1$ is a hypergeometric function of the first kind, and 
\beqa
\label{cero_2}
\langle \Psi^{\downarrow}_0(t_f)|\Psi^{\uparrow}_0(t_f) \rangle&=&e^{i[G^\downarrow(t_f)-G^\uparrow(t_f)]} e^{-\frac{m\omega}{\hbar}[(\delta \alpha_\epsilon)_{t_f}^2+(\delta \alpha_c)_{t_f}^2]}
\nonumber\\
&\times& e^{\frac{m}{\hbar \omega}(\omega \delta \alpha_c + i \dot \delta \alpha_\epsilon)_{t_f}^2}
\eeqa 
is a common factor in all terms. 
We also find (as before $n'\geqslant n$),  
\beqa
\label{np_n}
&&\hspace*{-.7cm}\langle \Psi_{n'}^{\downarrow}(t_f)|\Psi_{n}^{\uparrow}(t_f)\rangle\! =
\nonumber\\
&&\hspace{-.7cm} \langle \Psi^{\downarrow}_0(t_f)|\Psi^{\uparrow}_0(t_f) \rangle ({-\delta Z_\epsilon(t_f)})^{(n'-n)} \frac{(-2)^{(n'-n)}}{(n'-n)!} \sqrt{\frac{n'!}{n!}} \nonumber \\
&&\hspace{-.7cm}\times_1F_1 \left\{-n; n'-n+1; 4|{\delta Z_\epsilon}(t_f)|^2 \right \}.
\eeqa

The phase difference is
\beqa
\label{g_dif_phases_2}
G^\downarrow(t_f)-G^\uparrow(t_f)&=&\frac{1}{\hbar} \left [ -m(\dot \delta \alpha_\epsilon \delta \alpha_c+ \dot \delta \alpha_c \delta \alpha_\epsilon)_{t_f}\right.
\nonumber\\
&&\hspace{-3cm}+\left.2c\int_0^{t_f} \alpha dt + c \int_0^{t_f} \delta \alpha_\epsilon dt+\epsilon \int_0^{t_f} \delta \alpha_c dt \right ].
\eeqa
If there is no error in the driving force, $\epsilon=0$, the phase difference is given by Eq. (\ref{pd0}). 
Notice also that, according to  Eq. (\ref{de_dc}),  for the special times $t_f=n(2\pi/\omega)$  the terms of type $(\dot \delta \alpha_\epsilon \delta \alpha_c)_{t_f}$ and 
$(\dot \delta \alpha_c \delta \alpha_\epsilon)_{t_f}$ in Eqs. (\ref{cero_2}) and (\ref{g_dif_phases_2}) vanish. 

If the perturbations $c$ and $\epsilon$ are of the same order, the phase difference is in first order 
\beq
\label{dif_phases_2}
G^\downarrow(t_f)-G^\uparrow(t_f)=\frac{2c}{\hbar} \int_0^{t_f} \alpha(t) dt,
\eeq
as in Eq. (\ref{pd0}), i.e., there is no first order contribution of the error.
If the initial state is the ground state, the corrections to phase and modulus are only 
of second order.   

For a general state, to calculate the overlap we have to sum over all states 
(the overlap may be calculated in any picture, in particular in the rotating picture) 
\beq
\langle \psi_{\downarrow}(t_f)|\psi_{\uparrow}(t_f) \rangle =  \sum_{n,n'} (c_n)^* c_{n'} \langle \Psi_{n}^{\downarrow}(t_f)|\Psi_{n'}^{\uparrow}(t_f) \rangle.
\eeq
The $c_n$ do not carry a spin up/down superscript because the up and down states are equal at the initial time. 
Taking into account Eq. (\ref{overlap_epsilon}), first order  corrections to the zeroth order overlap must come from 
``nearest neighbour'' values of the integers, 
\beqa
\langle \psi_{\downarrow}(t_f)|\psi_{\uparrow}(t_f) \rangle &=& \sum_n (c_n)^* c_{n} \langle \Psi_{n}^{\downarrow}(t_f)|\Psi_{n}^{\uparrow}(t_f) \rangle
\nonumber\\
&&\hspace{-2cm}+(c_n)^* c_{n+1} \langle \Psi_{n}^{\downarrow}(t_f)|\Psi_{n+1}^{\uparrow}(t_f) \rangle
\nonumber \\
&&\hspace{-2cm}+(c_{n+1})^* c_{n} \langle \Psi_{n+1}^{\downarrow}(t_f)|\Psi_{n}^{\uparrow}(t_f) \rangle+...
\eeqa
The first order correction to $\langle \Psi^{\downarrow}_0(t_f)|\Psi^{\uparrow}_0(t_f) \rangle$
is proportional to $i(\dot{\delta}\alpha_\epsilon)_{t_f}\langle \Psi^{\downarrow}_0(t_f)|\Psi^{\uparrow}_0(t_f) \rangle$. 
This structure implies that there are no first order corrections to the modulus due to the error whereas in principle there 
is a first order correction to the phase. To make it vanish it is enough to take for the final time one of the periods, 
namely $t_f=n 2\pi/\omega$, see Eq. (\ref{de_dc}).        

 
%
%

%
%
%
%
\subsection{Only one branch  perturbed by $c$}
Now we study the scenario in which the homogeneous, small constant force $c$ acts only on one of the spin states. We consider 
the force $f_\alpha+c$ acting on the spin state $|\uparrow\rangle$ while the other spin state $|\downarrow\rangle$ is only subjected to the force $-f_\alpha$.
Substituting $\delta y^\downarrow_c \rightarrow 0$ and $\delta \alpha_\epsilon \rightarrow 0$, the phase difference is 
\beqa
G^\downarrow(t_f)-G^\uparrow(t_f)&=&\frac{1}{2\hbar} \left [ -m (\dot \delta \alpha_c \delta \alpha_c)_{t_f}\right.
\nonumber\\
&+&\left.2c\int_0^{t_f} \alpha dt + c \int_0^{t_f} \delta \alpha dt \right ].
\eeqa
The overlap for $n'\geqslant n$ can be written as
\beqa
\label{overlap_onemode}
&&\hspace*{-.5cm}\langle \Psi_{n}^{\downarrow}(t_f)|\Psi_{n'}^{\uparrow}(t_f)\rangle
= \langle \Psi^{\downarrow}_0(t_f)|\Psi^{\uparrow}_0(t_f) \rangle (\delta Z_c(t_f)^*)^{(n'-n)}
\nonumber \\
&&\hspace{-.5cm}\times\sum_{k=0}^\infty (-1)^{(k+n'\!-n)}\! \!\frac{n!}{(n-k)!} \sqrt{\frac{n'!}{n!}} \frac{[|\delta Z_c(t_f)|^2]^k}{k! (k+n'-n)!}
\nonumber\\
&=& \langle \Psi^{\downarrow}_0(t_f)|\Psi^{\uparrow}_0(t_f) \rangle (\delta Z_c(t_f)^*)^{(n'-n)} \frac{(-1)^{(n'-n)}}{(n'-n)!} \sqrt{\frac{n'!}{n!}}
\nonumber \\
&&\hspace{-.5cm}\times\; _1F_1 \left[-n; n'-n+1;|\delta Z_c(t_f)|^2 \right]\!,
\eeqa
where now  
\beqa
\langle \Psi^{\downarrow}_0(t_f)|\Psi^{\uparrow}_0(t_f) \rangle&=&e^{i[G^\downarrow(t_f)-G^\uparrow(t_f)]} e^{-\frac{m\omega}{2\hbar}(\delta \alpha_c)_{t_f}^2}
\nonumber\\
&\times& e^{\frac{m}{4 \hbar \omega}(\omega \delta \alpha_c + i \dot \delta \alpha_c)_{t_f}^2},
\eeqa
whereas 
\beqa
\label{overlap_onemode}
&&\hspace*{-.5cm}\langle \Psi_{n'}^{\downarrow}(t_f)|\Psi_{n}^{\uparrow}(t_f)\rangle
\nonumber\\
&=& \langle \Psi^{\downarrow}_0(t_f)|\Psi^{\uparrow}_0(t_f) \rangle (-\delta Z_c(t_f))^{(n'-n)} \frac{(-1)^{(n'-n)}}{(n'-n)!} \sqrt{\frac{n'!}{n!}}
\nonumber \\
&&\hspace{-.5cm}\times\; _1F_1 \left[-n; n'-n+1;|\delta Z_c(t_f)|^2 \right]\!.
\eeqa
%
%
%
The analysis of the corrections of 
$
\langle \psi_{\downarrow}(t_f)|\psi_{\uparrow}(t_f) \rangle =e^{\frac{ic}{\hbar} \int_0^{t_f} \alpha dt},
$
is parallel to the one in the previous subsection. 
For the ground state, corrections of phase and modulus are second order in $c$, while for an arbitrary state, 
the correction to the modulus  is of second order and the first order correction to the phase 
can be made zero choosing the final time to be an integer of the period $2\pi/\omega$.  
%
%

%
%
%
%
%
%
%
\section{Designing forces by inverse engineering techniques.}
In order to inverse engineer the driving force we must first set a trajectory $\alpha(t)$,  
a particular solution of the Newton equation (\ref{NE}) which satisfies the boundary conditions (\ref{bc_y}).

\begin{figure}[htbp]
{\includegraphics[width=85mm]{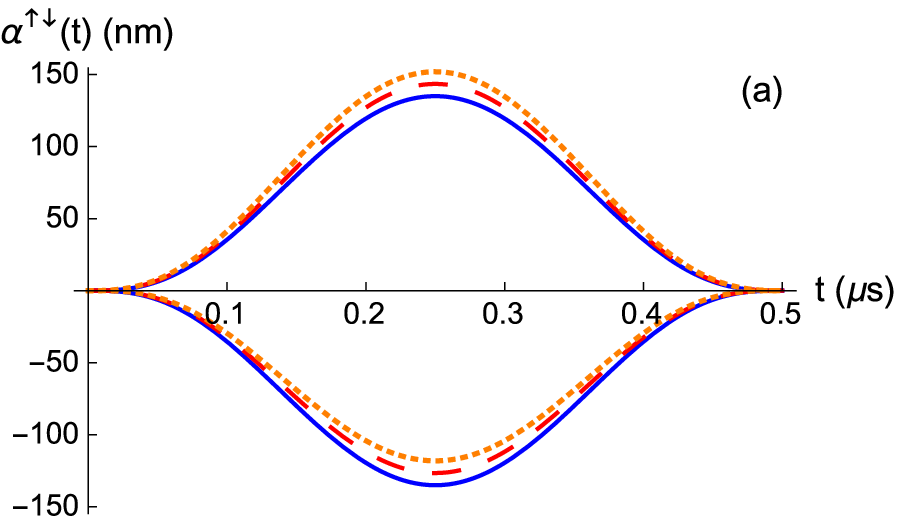}}
{\includegraphics[width=85mm]{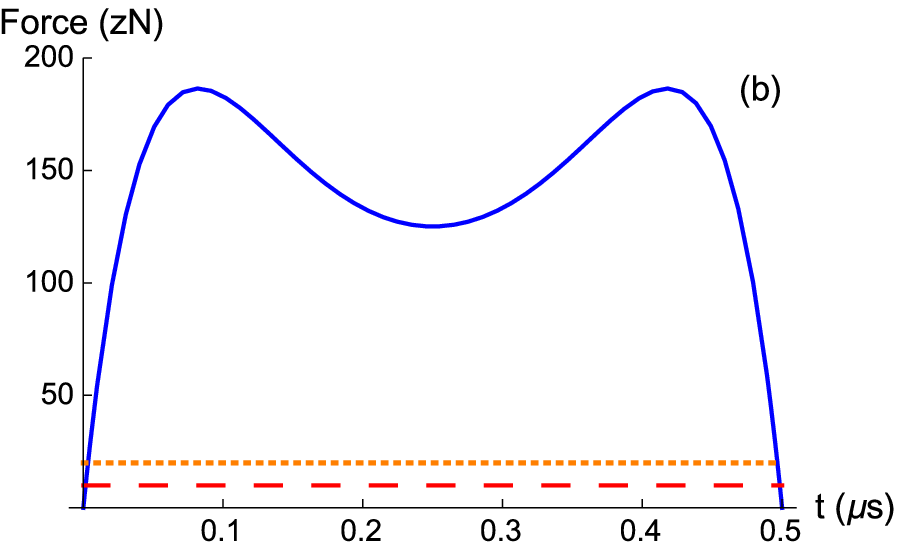}}
\caption{(Color online) (a) Unperturbed trajectory $\alpha_{A}(t)$ (blue solid line), and trajectories perturbed by a force $c=10$ zN (red dashed line), and $c=20$ zN (orange dotted line), at $t_{f}=0.5$ $\mu$s. We have selected $\alpha_{A}(t_{f}/2)=135$ nm, so that sensitivity is $\int_{0}^{t_{f}}\alpha_{A}dt=30.9$ nm$\times \mu$s.
(b) Force for the unperturbed trajectory (blue solid line), and additional forces $c=10$ zN (red dashed line), and $c=20$ zN (orange dotted). ${\omega}/({2\pi})=2$ MHz.}
\label{fig:alfaAforceA}
\end{figure}

We have used sixth order polynomials to design a first type of trajectory
\begin{equation}
\label{alphaA}
\alpha_{A}(t)=\sum_{j=0}^{6} b_{j} \left(\frac{t}{t_{f}}\right)^{j}. 
\end{equation}
After selecting the final time $t_{f}$, and applying the six boundary conditions in Eq. (\ref{bc_y})
we still have one free parameter left. To fix it  we impose a value to the trajectory at $t_f/2$, 
\begin{equation}
\label{ExtraCondition}
\alpha_{A}\left(\frac{t_{f}}{2}\right)=M,
\end{equation}
which is the maximum displacement of the trajectory (see Fig. \ref{fig:alfaAforceA}). We may increase the sensitivity  
$S=\int_0^{t_f} \alpha dt$ by selecting a larger $t_{f}$ and/or a larger value of $M$.  
The corresponding driving force is calculated from the Newton equation (\ref{NE_lr_2}), 
with ${\omega}/({2\pi})=2$ MHz and the mass of a $^{9}Be^{+}$ ion.

%
%
%
%

\begin{figure}[htbp]
{\includegraphics[width=85mm]{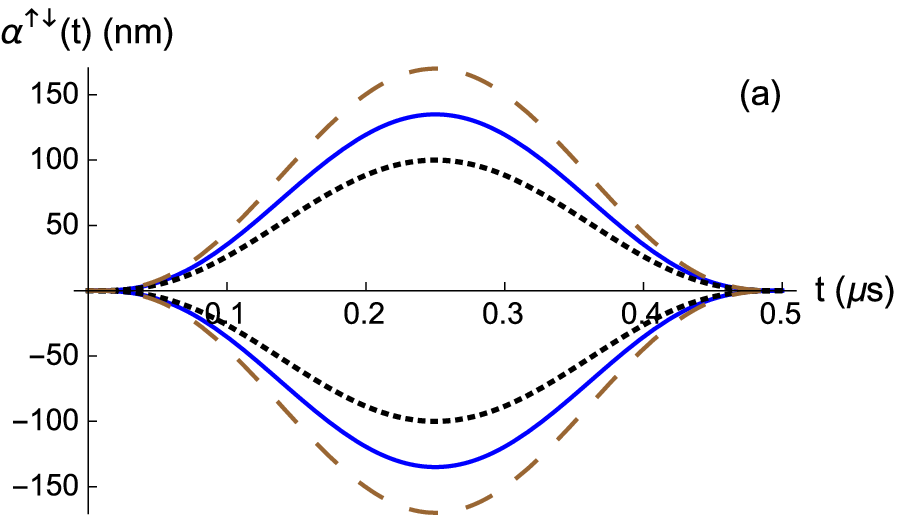}}
{\includegraphics[width=85mm]{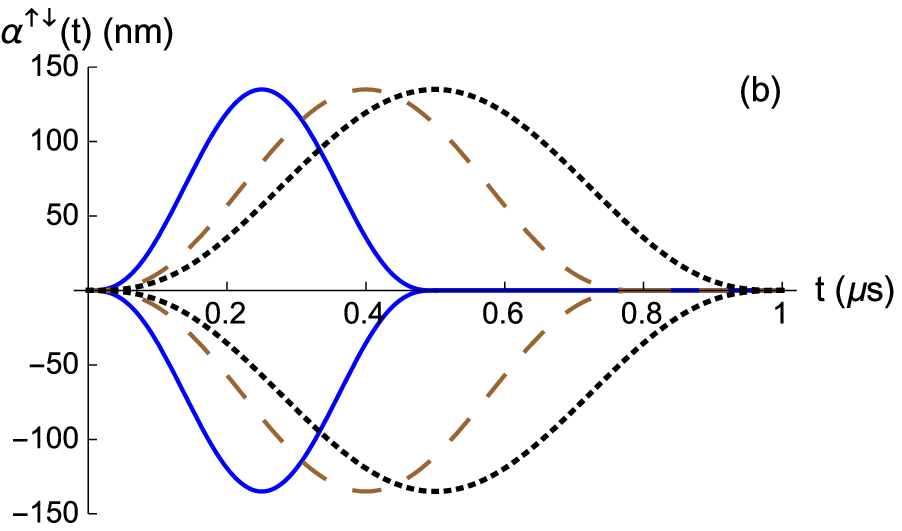}}
\caption{(Color online) (a) Trajectories for $t_{f}=0.5$ $\mu$s for different values of the extra condition (\ref{ExtraCondition}) $\alpha_{A}(t_{f}/2)=100$ nm (black dotted line), $\alpha_{A}(t_{f}/2)=135$ nm (blue solid line) and $\alpha_{A}(t_{f}/2)=150$ nm (brown dashed line). (b) Trajectories with the same value of the extra condition (\ref{ExtraCondition}) $\alpha_{A}(t_{f}/2)=135$ nm for final times $t_{f}=0.5$ $\mu$s (blue solid line), $t_{f}=0.8$ $\mu$s (brown dashed line) and $t_{f}=1$ $\mu$s (black dotted line).}
\label{fig:sensitivity}
\end{figure}

In Fig. \ref{fig:alfaAforceA} we plot a trajectory $\alpha_{A}(t)$, the driving force $f_{\alpha,A}(t)$, and the trajectories perturbed by two values of the homogeneous force $c$. The perturbation to the trajectory $\alpha_{A}(t)$ due to $c$ is given by Eq. (\ref{de_dc}). 
In Fig. \ref{fig:sensitivity} we plot other $\alpha_{A}(t)$ trajectories.  In Fig. \ref{fig:sensitivity} (a) we vary $M$  in Eq. (\ref{ExtraCondition}) for a given $t_f$, whereas in Fig. \ref{fig:sensitivity} (b), $M$ is fixed and different final times are used.    

%
%
%
%

\begin{figure}[htbp]
\includegraphics[width=85mm]{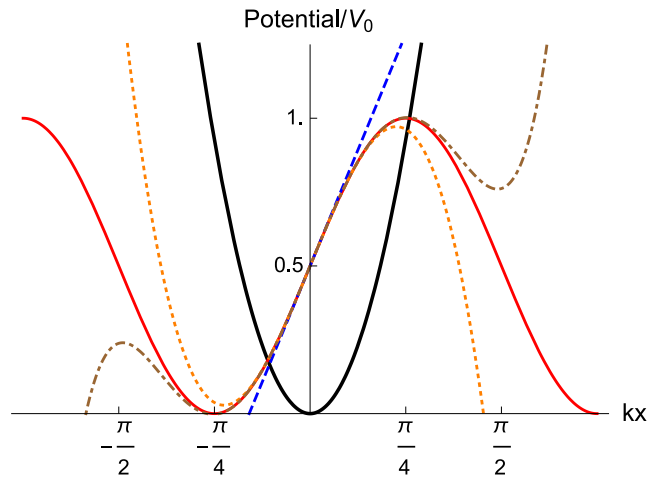}
\caption{Schematic picture of the trapping potential (black solid line), real optical potential (red solid line) given by $V_{0}\sin^{2}(kx+\frac{\pi}{4})$, 
and linear (blue dashed line), third order (orange dotted line) and fifth order (brown dot-dashed line) approximations to the optical lattice using the Taylor series around $x=0$. 
}
\label{fig:OpticalLattice}
\end{figure}

So far, we have considered a homogeneous force, see Eq. (\ref{Hamiltonian}), but if the optical force is implemented by an optical lattice, it is not really homogeneous.  Figure  \ref{fig:OpticalLattice} depicts a schematic picture of a harmonic trapping potential, the potential generated via an optical lattice (red) $V_0\sin^{2}(kx+\frac{\pi}{4})$, and three different approximations keeping the  linear, cubic, and quintic terms in the Taylor series around $x=0$. To mitigate the effect 
of deviations from the homogeneous force regime, we can design different trajectories that, for a given value of the sensitivity $\int_0^{t_f} \alpha dt$, reduce or even minimize the maximal deviation. A way to estimate the importance of the cubic  term is to calculate $\int_{0}^{t_{f}}\alpha^{3}dt$. The smaller such a value is, the better the linear approximation. 

%
%
%
%

\begin{figure*}[htbp]
\centering
{\includegraphics[width=85mm]{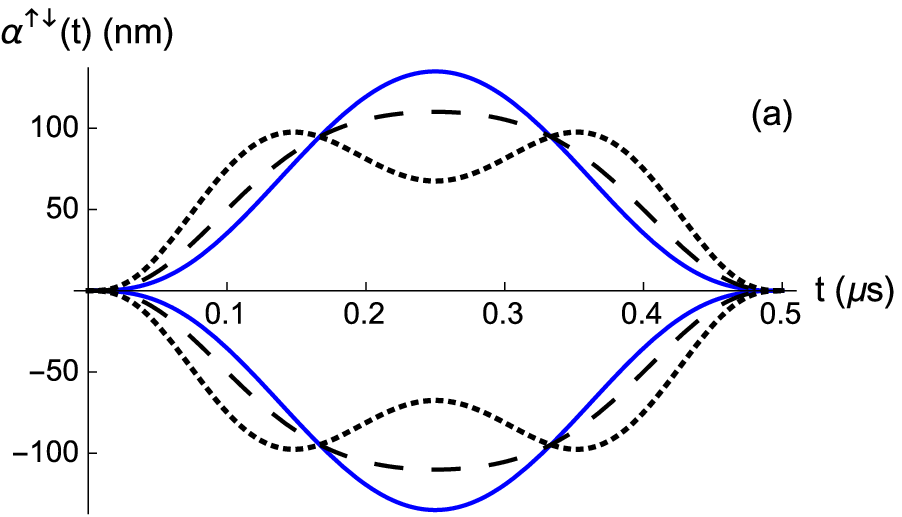}}
{\includegraphics[width=85mm]{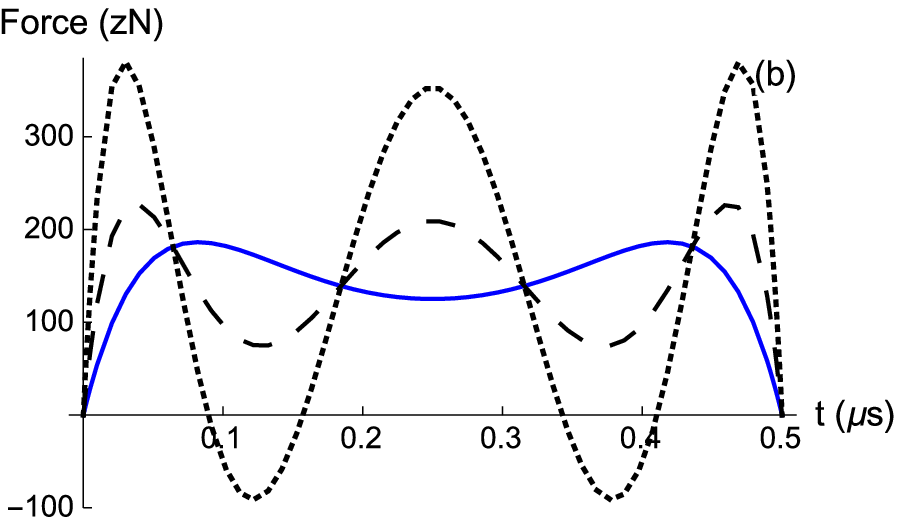}}
{\includegraphics[width=85mm]{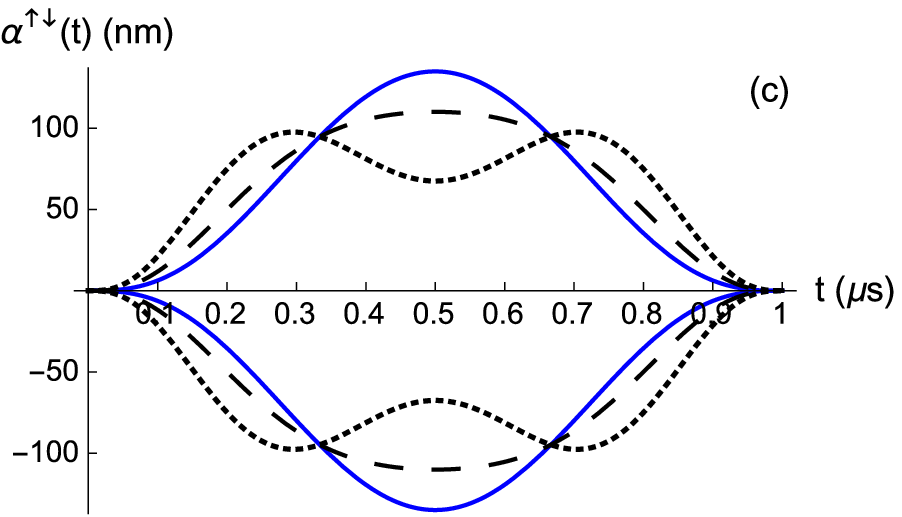}}
{\includegraphics[width=85mm]{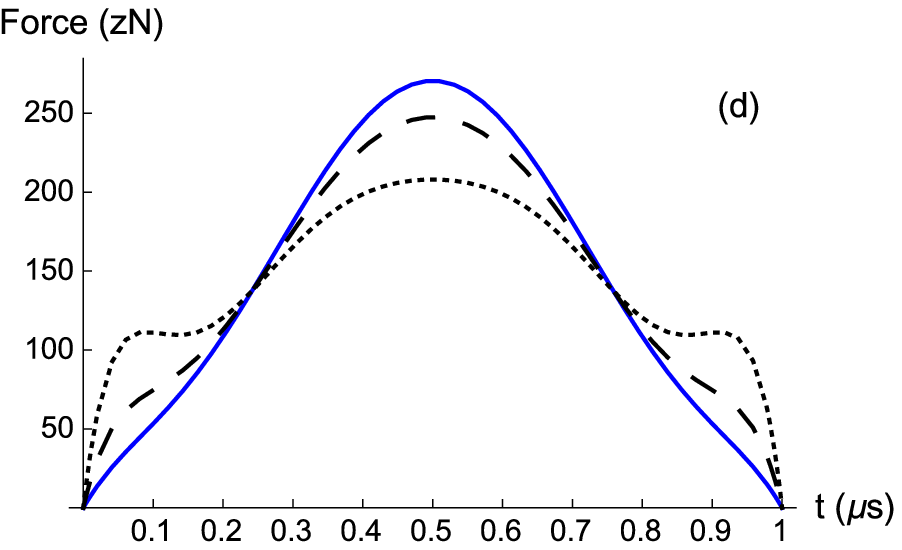}}
\caption{(Color online) Trajectories and their corresponding forces. (a,b): $t_f= 0.5$ $\mu$s; (c,d): $t_f=1$ $\mu$s.  
$\alpha_{A}(t)$ and $f_{\alpha, A}(t)$ (blue solid lines);  $\alpha_{B}(t)$ and $f_{\alpha, B}(t)$ (black lines, dashed for $v=50$ nm, dotted for $v_{opt}=75$ nm). At $t_{f}= (0.5, 1)$ $\mu$s, $\int_{0}^{t_{f}}\alpha_{A}^{3}dt=(3.49\times 10^{5}, 6.98\times 10^{5})$ nm$^{3}\times \mu$s, whereas $\int_{0}^{t_{f}}\alpha_{B}^{3}dt=(2.05\times 10^{5}, 4.10\times 10^{5})$ nm$^{3}\times \mu$s for $v_{opt}$. }
\label{fig:comparation}
\end{figure*}

Thus, consider a new type of trajectories, $\alpha_{B}(t)$, 
%
\begin{equation}
\label{alphaB}
\alpha_{B}(t)=\sum_{j=0}^{8} a_{j}  \left(\frac{t}{t_{f}}\right)^{j}. 
\end{equation}
After selecting $t_{f}$ and applying the six boundary conditions in Eqs. (\ref{bc_y}),
we still have three free parameters left. To fix them we impose the sensitivity with $\alpha_{B}$-trajectories to be the same than the 
sensitivity with $\alpha_{A}$-trajectories. We also impose a value, $v$,  at $t=\frac{t_{f}}{5}$ and $t=\frac{4t_{f}}{5}$,   
\begin{eqnarray}
\label{eq:ThreeExtraConditions}
\int_{0}^{t_{f}}\alpha_{B}(t)dt&=&\int_{0}^{t_{f}}\alpha_{A}(t)dt, 
\nonumber \\
\alpha_{B}\left(\frac{t_{f}}{5}\right)&=&\alpha_{B}\left(\frac{4t_{f}}{5}\right)=v.
\end{eqnarray}
In Fig. \ref{fig:comparation} we plot the  trajectories  $\alpha_{A}(t), \alpha_{B}(t)$ and their corresponding forces $f_{\alpha, A}(t), f_{\alpha, B}(t)$, for two final times. At $t_{f}=0.5$ $\mu$s the $\alpha_{B}$-trajectories 
imply higher forces. Furthermore, for $v_{opt}$ (value at which the importance of the third order term in the optical potential is minimum) abrupt forces are required, reaching even negative values. However at moderately larger times, specifically for $t_{f}=1$ $\mu$s,  
$\alpha_{B}$-trajectories require smaller, softer forces.

%
%
%
%

\begin{figure}[htbp]
\includegraphics[width=80mm]{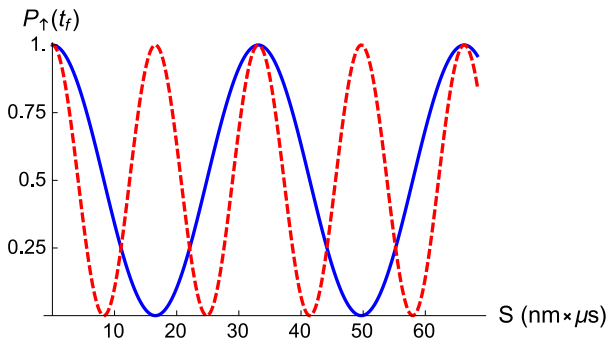}
\caption{Spin-up state population after a $\pi/2$-pulse at final time $t_{f}=0.5$ $\mu$s as a function of the sensitivity of the interferometer $S=\int_0^{t_f} \alpha dt$, which is varied changing the maximum displacement 
$M$, Eq. (\ref{ExtraCondition}), using $\alpha_A$ trajectories 
(\ref{alphaA}). $c=10$ zN (blue solid line) and $c=20$ zN (red dashed line).}
\label{fig:populations}
\end{figure}

\textbf{Measuring $c$ by measuring populations}:
The phase measurement is done through measurements of the populations for spin up or spin down. 
Using Eqs. (\ref{populations_2}) and (\ref{o}), and assuming modulus one in Eq. (\ref{o}), we have
\begin{eqnarray}
\label{measurement}
P_{\uparrow}(t_{f})&=&\frac{1}{2}+\frac{1}{2}\cos\Delta\phi(t_{f}),
\nonumber \\
P_{\downarrow}(t_{f})&=&\frac{1}{2}-\frac{1}{2}\cos\Delta\phi(t_{f}),
\end{eqnarray}
where $\Delta\phi(t_{f})=G_\alpha^{\downarrow}(t_{f})-G_\alpha^{\uparrow}(t_{f})=\frac{2c}{\hbar}\int_{0}^{t_{f}}\alpha(t) dt=\frac{2c}{\hbar} S$ or 
\beq
c=\frac{\hbar\Delta\phi(t_{f})}{2\int_0^{t_f} \alpha dt}=\frac{\hbar\Delta\phi(t_{f})}{2S}.
\eeq
However, as the populations (\ref{measurement}) 
are periodically oscillating functions of $\Delta\phi(t_f)$, a single value of the population corresponds to an infinite number of phases $\Delta\phi(t_{f})$.   
To avoid such ambiguity and extract $c$ from the ``real'' phase, we may use different values for the sensitivity of the interferometer $S$ (designing different unperturbed trajectories) and measure the populations for all of these values, so that we can plot the population of each state as a function of the sensitivity of the interferometer. 
$c$ can be found from the oscillation ``period'' $\pi \hbar/c$ of the populations with respect to $S$, see  Fig. \ref{fig:populations}.  

%
%
%
%

\section{Discussion}
We have presented the theory to perform driven interferometry to a trapped ion separating the wavefunction 
branches with controllable spin-dependent homogeneous forces. Specifically we have considered the measurement of an 
unknown homogeneous force when the ion is trapped in a harmonic potential. 
Invariant-based engineering has been proposed to design the spin-dependent forces and 
to control the sensitivity, the process time, or to minimize the effect of anharmonicities. 
The control of both the final time and the sensitivity allows us to work in a diabatic regime to avoid decoherence. 
The measured phase is shown to be robust with respect to constant errors in the implementation of the control forces.
As for phase gates based on similar principles \cite{Palmero2017}, it is also independent of the motional state within the
assumed harmonic trap approximation.    
  
Several extensions of this work are possible, e.g. to implement driving forces that minimize the effect of different noises
\cite{Ruschhaupt2012}. Specific, setting dependent forms or values for $x_0(t)$ may be considered, such as for example 
an oscillating function. 
It is worth noting that invariants are also explicitly known for more complicated
configurations \cite{Torrontegui2011},
in particular
for Hamiltonians with rigidly moving potentials 
complemented by a time-dependent linear term that compensates the inertial forces. 
%
%
This setting may be suitable for optical lattices. 
Finally, we intend to extend these results to oscillating forces. 
\acknowledgments{We are grateful to D. Leibfried and J. Bollinger for their suggestions and comments. 
This work was supported by the Basque Country Government (Grant No. IT986-16), and MINECO/FEDER, UE (Grant No. FIS2015-67161- P).}

\appendix
\section{General solution for the forced harmonic oscillator\label{apa}}
In this appendix we assume a forced harmonic oscillator with Hamiltonian (\ref{Hamiltonian}).
To describe a general trajectory for the Newton equation (\ref{NE}),
it is useful to define dimensionless positions and momenta as 
\beqa
\label{cuadrature}
Y=\sqrt{\frac{m\omega}{2 \hbar}}y,\;\;\;
P=\sqrt{\frac{1}{2\hbar m\omega}}p,
\eeqa
as well as complex-plane combinations $Z=Y+iP$.
The general solution of the position and momentum of a classical particle,
or the corresponding expectation values for any quantum state, is compactly given in complex form as
\beqa
\label{solution_xp}
Z(t)&=&e^{-i\omega t} \left \{  Z(0) + \frac{i}{\sqrt{2 \hbar m \omega}} \int_0^t d\tau e^{i\omega \tau} F \right \}
\nonumber \\
&=& \tilde Z +Z_0,
\eeqa
where 
\beqa
\tilde Z &\equiv& e^{-i \omega t} Z(0),
\\
\label{position}
Z_0&\equiv& \sqrt{\frac{m\omega}{2 \hbar}} y_0 + i \sqrt{\frac{m}{2 \hbar \omega }} \dot y_0,
\eeqa
and $y_0$ is a particular solution satisfying $y_0(0)=\dot y_0(0)=0$. For an $F=f_\alpha$ such that $y_0(t)=\alpha(t)$,
and thus $Z_0=Z_{\alpha}$, the boundary conditions at $t_f$ are satisfied as well in the particular solution
[see Eq. (\ref{bc_y})]. 

By separating into real and imaginary parts, it can be seen that 
\beq
\Re{\rm e} (\tilde{Z}) \frac{1}{\sqrt{2\hbar m \omega}}F=\frac{\partial \Im{\rm m}(Z_0 \tilde{Z}^*)}{\partial t},
\label{re-im}
\eeq
so that
\beq 
\int_0^{t_f} dt\, \Re{\rm e} (\tilde{Z}) f_\alpha=0, 
\label{zero}
\eeq
since $Z_{\alpha}(0,t_f)=0$.
This result is used repeatedly, e. g. in Sec. \ref{4A} or in Appendix B.      

\section{Relation between total, dynamical, and geometric phases\label{apb}}
A forced harmonic oscillator (\ref{Hamiltonian}) with force $F(t)$ leads to a common final phase 
$\phi=-G(t_f)$,  see Eq. (\ref{IPphases}), irrespective of the initial state in the rotating frame  
with dynamical equation 
\beq
i \hbar \frac{\partial \psi_I}{\partial t}=V_I \psi_I, 
\eeq
where $V_I=-F e^{i H_{0}t/\hbar}xe^{-i H_{0}t/\hbar}$. 
The total phase may be split into dynamical and geometric contributions. 
The dynamical phase is defined by 
\beqa
\label{d_phase}
\phi_d&=&-\frac{1}{\hbar}\int_0^{t_f} dt \langle \psi_I(t)|V_I(t)|\psi_I(t)\rangle
 \nonumber \\
&=&-\frac{1}{\hbar}\int_0^{t_f} dt \langle \psi(t)|V(t)|\psi(t)\rangle
\nonumber \\
&=&\frac{1}{\hbar}\int_0^{t_f} dt F(t) \langle x(t) \rangle,
\eeqa
where $V=-F(t) x$. 
From Ehrenfest's theorem, the expectation value of $x$ corresponds to a classical trajectory, i.e.,
to a solution of Eq. (\ref{NE}), but not necessarily to the special solutions $\alpha(t)$.
Using the phase-space trajectory in the rotating frame $Z_r=e^{i \omega t} Z=Y_r+iP_r$, 
see Appendix A,  we may write $F \langle x \rangle /\hbar=F \sqrt{\frac{2}{\hbar m \omega}}\Re{\rm e}(Z)=2\Im{\rm m}\left(\frac{dZ_r}{dt}Z_r^*\right)=4d\mathcal{A}/dt$, where $d\mathcal{A}$ is the differential of area swept in the rotating phase space, $d\mathcal{A}/dt=\frac{Y_r}{2}\frac{d P_r}{dt}-\frac{P_r}{2}\frac{d Y_r}{dt}$. Thus Eq. (\ref{d_phase}) becomes 
\beq
\phi_d=4\mathcal{A}.
\eeq

Using Eq. (\ref{re-im}) we rewrite Eq. (\ref{d_phase}) as
\beq
\label{general_phid}
\phi_d=\frac{1}{\hbar}\int_0^{t_f}dt \left [ y_0+\sqrt{\frac{2 \hbar}{m \omega}}
\Re(\tilde{Z}) \right ] F.
\eeq
A particular case of interest is $y_0=\alpha$, $F=f_\alpha$. Due to Eq. (\ref{zero}) the dynamical phase 
becomes $\frac{1}{\hbar}\int_0^{t_f}dt \alpha=2\phi$. Thus $\phi_g=-\phi_d=\phi$.

To apply these results we start from the Hamiltonian (\ref{DidiH_A}), 
%
%
and, considering $x_0$ constant, we apply the change $\tilde{x}(t)=x-x_0$ 
(different from the one in Eq. (\ref{DidiH_A_new})) to rewrite the Hamiltonian as
\begin{eqnarray}
\label{H_af_new}
H&=&\frac{p^{2}}{2m}+\frac{1}{2}m\omega^{2}\tilde{x}^{2}-\left[\tilde{c}+f(t,\sigma^{z}) \right]\tilde{x}
\nonumber \\
&-&\tilde{c}x_0-\frac{1}{2}m\omega^2x_0^2, 
\end{eqnarray}
where $\tilde{c}=c-m\omega^2 x_0$. 
%
%
The last two terms are and constant and spin-independent so we may ignore them to 
set Hamiltonians of the form (\ref{Hamiltonian}).

We consider now that the effective forces acting on both branches are $f_\alpha+\tilde{c}$ for $\sigma^z=1$ and $-f_\alpha+\tilde{c}$ for $\sigma^z=-1$. 
%
%
Following the same procedure as in Sec. \ref{4A}, with $\epsilon=0$, and $c\to\tilde{c}$, the phase difference at final time is
found to be 
\beq
\label{deltaphi_d}
\Delta \phi(t_{f})=\phi_{\uparrow}-\phi_{\downarrow}=\frac{2\tilde{c}}{\hbar}\int_{0}^{t_{f}} \alpha(t)dt,
\eeq
in agreement with Eq. (\ref{Didi_PhaseDifferenceB_new}). 

%
%
%
%
%
%
%
%
%
Substituting in Eq. (\ref{general_phid}) for the spin-up branch ($F=f_\alpha+\tilde{c}$) 
and for a general trajectory (i.e., with an arbitrary initial condition) we find  that
\beqa
\phi_d^\uparrow&=&\frac{1}{\hbar}\int_0^{t_f}\!\! (\alpha+\delta\alpha_{\tilde{c}})(f_\alpha+\tilde{c})dt
\nonumber \\
&+&
\frac{1}{\hbar}\sqrt{\frac{2\hbar}{m\omega}}\int_0^{t_f}\!\! \Re e \tilde{Z}  (f_\alpha+\tilde{c}) dt
\eeqa
with corresponding results for $F=-f_\alpha+\tilde{c}$.   
Using the explicit form of $\delta \alpha_{\tilde{c}}$, see Eq. (\ref{de_dc}), Eq. (\ref{integr}) and Eq. (\ref{zero}), 
and subtracting, 
\beq
\label{phid_area}
\Delta \phi_d=\phi_d^\uparrow-\phi_d^\downarrow= 
\frac{4\tilde{c}}{\hbar}\int_0^{t_f} \alpha(t) dt =4\Delta {\mathcal A},
\eeq
which does not depend on the specific trajectory (value of $Z(0)$).   
Comparing with (\ref{deltaphi_d}) we also have $\Delta \phi_g=-2\Delta \mathcal{A}$, and $\Delta \phi=2\Delta \mathcal{A}$. 
%
%
As mentioned in Sec. \ref{br}, if $x_0=\frac{c}{m \omega^2}$, then $\tilde{c}=0$, 
two branch areas are equal [see Fig. \ref{fig:scheme_areas} (b)], and the phase differentials vanish.

\label{Bibliography}
\bibliographystyle{unsrt}
\bibliography{Bibliography}
\end{document}